# Three-Photon Saturable Absorption in Atomically Thin Phlogopite

Nabarun Mandal[1#], Sagnik Chakraborty[1#], Ranjeet Singh[2], Jhionathan de Lima[4], Saswata Goswami[1], Binay Bhushan[3], Rahul Rao[5], Nicholas R. Glavin[5], Ajit K. Roy[5], Vidya Kochat[1], Cristiano Francisco Woellner[4*], Prasanta Kumar Datta[2*], and Chandra Sekhar Tiwary[1*]

[1]*School of Nanoscience and Technology, Indian Institute of Technology Kharagpur, 721302, India.*

[2]*Department of Physics, Indian Institute of Technology, Kharagpur, 721302, India.*

[3]*Department of Physics, Birla Institute of Technology, Mesra-Patna Campus, Patna, Bihar, 800014, India*

[4]*Department of Physics, Federal University of Parana UFPR Curitiba-PR, 81531-980, Brazil*

[5]*Materials and Manufacturing Directorate, Air Force Research Laboratory, Wright Patterson AFB, Ohio 45433-7718, United States.*

[#]*Contributed equally*

**Corresponding authors, E-mail: woellner@ufpr.br, pkdatta@phy.iitkgp.ac.in, chandra.tiwary@metal.iitkgp.ac.in*

## Abstract

Two-dimensional (2D) materials, due to their remarkable physical and chemical properties, hold significant potential for future optical and electrical applications. In this study, the synthesis of 2D phlogopite (magnesium-rich mica) via liquid-phase exfoliation (LPE) is reported using an efficient and scalable procedure. XRD structural analysis revealed a preferential orientation along the (033) plane, whereas AFM and SEM demonstrated a nanoscale thickness and homogeneous morphology. Optical characterization by UV-Vis and Raman spectroscopy reveals tunable band gaps up to 4.52 eV in exfoliated 2D phlogopite and distinct vibrational modes indicative of structural evolution. At intense laser conditions, three-photon saturable absorption (3PSA) behavior was detected by light-modulated electrical property studies, which emphasize the potential for optical limiting, switching, and mode-locking applications. The present investigation indicates 2D phlogopite is a versatile material for next-generation optoelectronic devices of high-intensity light modulation.

## 1. Introduction

Two-dimensional materials (2D materials) are a unique class of nanomaterials characterized by their ultra-thin structure, like a single or a few layers of atoms, varying significantly from their bulk form counterparts, mainly due to quantum confinement, interface/surface interfaces, altered electronic structure, and the material's atomic-level interfaces. [1-2] Such unique characteristics arise due to their reduced dimensionality, making them highly versatile and promising for a wide range of applications, including sensors, photodetectors, nonlinear optics, electronics, catalysts, and corrosion. [3-5]

After the discovery of graphene [6], some of the recent discoveries of 2D natural minerals are biotite [7], red coral [8], hematene (2D $\alpha$-$Fe_2O_3$) [9], ilmenene (2D $FeTiO_3$) [10], etc. Among them, phlogopite ($KMg_3AlSi_3O_{10}(F, OH)_2$) is the naturally occurring magnesium-rich mica that has been proven to be of interest due to its unique optical and electrical properties, as well as mechanical stability and resistance to high temperatures, making it applicable for use in extreme operational conditions. [11] It is most frequent in high-temperature metamorphic environments and at lower pressures, typically in connection with calc-silicate rocks, such as marbles and skarns. It forms by metamorphism of magnesium-rich carbonate rocks in silica-rich fluid environments and typically occurs along with minerals such as diopside and forsterite. [12] It is also very abundant in ultramafic igneous rocks, which include kimberlites [13] and lamproites. [14]

The interaction of light with a variety of materials often significantly alters their electrical properties and is of special importance in the fabrication of advanced optoelectronic devices. [15] Since this phenomenon involves, for instance, the photoexcitation of charge carriers, alteration in electronic band structure, and photothermal effects, light-induced modulation is useful in the fields of photodetectors, light-driven transistors, and energy conversion devices.

[16] Besides, phlogopite shows strong nonlinear optical properties, such as second and third harmonic generation, attributed to its layered silicate structure and strong anisotropy. [17] Phlogopite, with a wide band gap, shows great promise in optical switches, frequency converters, laser systems, photonic circuits, and even quantum communication because of the transparency that the wide band gap provides in the UV-visible spectrum range with high stability in both chemical and thermal grounds. [18] Light-driven modulation will help phlogopite assume a future role in photonic and optoelectronics technology. [19]

In this study, we fabricate 2D phlogopite by liquid phase exfoliation (LPE) from its bulk form and characterize its structural, optical, and electrical properties. The exfoliated 2D phlogopite is analysed using various techniques to understand its morphological characteristics, stability in suspension, and performance in optoelectronic applications. The phase and crystallographic information were obtained through X-ray Diffraction (XRD) Analysis using a PANalytical X'Pert diffractometer, which used Cu-kα radiation with a wavelength (λ) of 1.5406 Å and operating settings of 40 kV voltage and 40 mA current. Ultraviolet-visible (UV-Vis) spectroscopy (Analytical) was conducted to evaluate the linear optical properties within the wavelength range 200–1100 nm, and the optical band gap was determined using the tauc plot method. Surface morphology and layer thickness were analysed using tapping mode Atomic Force Microscopy (AFM) with a Bruker Dimension Icon AFM apparatus. Spectroscopy was carried out at ambient temperature by making use of the Raman technique with a WITec UHTS 300 VIS spectrometer (WITec GmbH, Germany) equipped with a 532 nm excitation laser. The general surface information was obtained by the Scanning Electron Microscope (SEM), and the elemental distribution analysis was carried out by the use of EDAX from the JEOL JXA-8530F system.

Here, we demonstrate the successful fabrication of 2D phlogopite by LPE from its bulk form and investigate its potential for optoelectronic applications. The exfoliated phlogopite is comprehensively characterized for its structural, optical, and electrical properties to explore its suitability for future advanced technologies. By utilizing various characterization techniques, we provide insights into the morphology, stability, and functionality of the 2D phlogopite material.

## 2. Experimental Section

### 2.1 Materials

All chemicals used in this experiment are purified analytical grade and were applied without further purification. The process of synthesis involved the use of both phlogopite and isopropyl alcohol (IPA), provided by Merck Chemical, LRL-LR grade.

### 2.2 Synthesis of 2D-Phlogopite

Phlogopite sheet (collected from Tip Top Mining District, Arizona, USA) was ground finely in a mortar and pestle. The obtained powder was kept at room temperature for later use. Subsequently, 50 mg of the above-obtained powdered material was dissolved in 50 mL of IPA. Since the sample-to-solution ratio was 1 mg/mL, the procedure above was followed. Then, 2D phlogopite was prepared via LPA, which was further continued for 10 hours. A relaxation time of 10 minutes was followed after every 20 minutes of exfoliation to maintain a temperature below 40°C. This method is relatively more efficient than other exfoliation techniques, such as the scotch-tape and bath sonication methods. After that, the dispersed solution was left to settle at ambient temperature for 48 hours. After allowing the non-exfoliating particles to settle, the solution that was dispersed and stable was left at room temperature and set aside for later use.

### 2.3 Experimental Setup

To perform the optical response experiment, the setup used included a 2-probe station for placing and preparing the sample, applying contacts on it, and a source for DC power for electrical measurement. Laser sources operating at 650 nm and 532 nm were used in addition for optical excitation. Precise alignment was ensured due to the mounting of the lasers on adjustable stands, and the measurement of the light incidence was done by Thorlabs s120c power sensor with a Thorlabs PM101 power meter. Under illuminated conditions, a constant voltage of 5 V was applied to the sample, and the corresponding current was measured as a function of time using precision probes. The resistance at each time point was calculated using Ohm's Law, enabling a time-resolved analysis of the sample's electrical behaviour under steady-state voltage and light exposure. The experiments were carried out under controlled conditions to minimize the interference of external sources.

## 3. Results and Discussions

### 3.1 Characterisation of 2D-Phlogopite

**Figure 1**a illustrates the UV-vis spectra of bulk phlogopite and its 2D counterpart. A specific absorption peak appears in the sample of 2D phlogopite, whereas it is not present in the spectrum of the bulk sample. Band gap values were evaluated from Tauc plots attained from absorption spectra. The band gap of the 2D phlogopite sample was determined to be 4.57 eV, whereas the bulk sample is 3.79 eV. The wide bandgap change with a variety of exfoliation durations can be attributed to the structural and electronic modifications resulting from exfoliation.

HRTEM images of 2D phlogopite are shown in **Figure 1**b(i)-(ii). With a d-spacing value of 3.37 Å, the inverse Fast Fourier Transform (FFT) lattice fringes are well matched to the (003) plane of phlogopite. This plane's maximal exfoliation is recognized in the XRD data as having a high intensity, as shown in **Figure 1**c.

The XRD spectra of bulk phlogopite and exfoliated 2D phlogopite are presented in **Figure 1**c. The obtained bulk spectrum is consistent with that of phlogopite and matches the monoclinic structure with reference to ICDD: 98-002-4163 and space group C1m1, 8. The strongest observed diffraction peak in 2D phlogopite corresponds to the (003) crystalline plane. It indicates that the (003) plane ($d$ = 3.37 Å) has a strong preferential orientation in the 2D phlogopite sample.

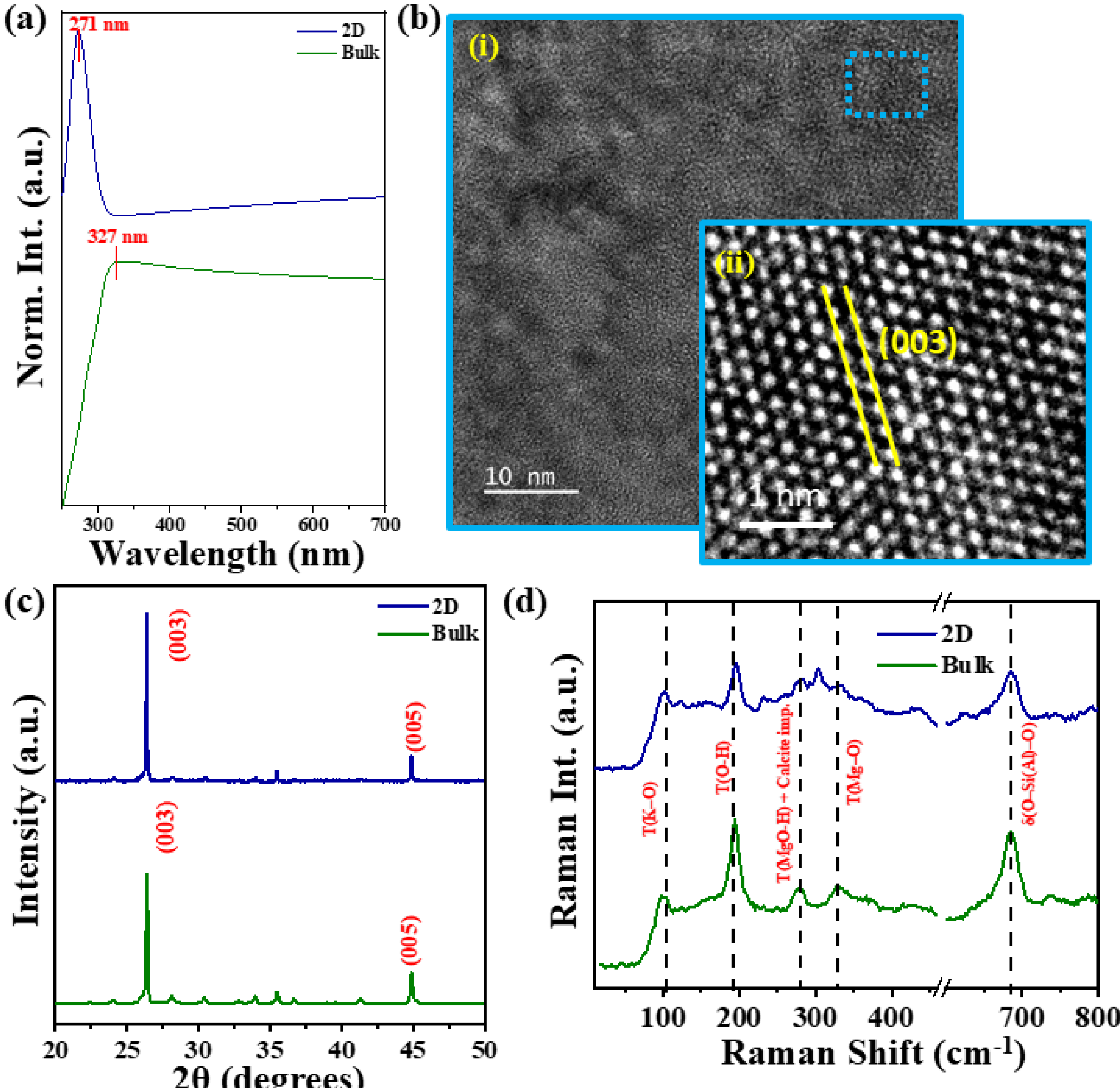

**Figure 1 (a)** UV-Vis spectra of bulk and 2D phlogopite, (b) HRTEM images of 2D phlogopite, (c) XRD spectra of bulk and 2D phlogopite, (d) Raman spectra of bulk and 2D phlogopite.

Raman spectra (**Figure 1**d) of phlogopite for different exfoliation times (bulk and 2D) show distinct changes in the vibrational modes, shedding light on the structural evolution of the material as it transitions from bulk to thinner layers. The peak near ~100 $cm^{-1}$, associated with the translational mode of K-O bonds, decreases in intensity progressively from the bulk to 2D, indicating a gradual weakening of interlayer interactions. A similar trend at ~200 $cm^{-1}$ peak, attributed to O-H vibrations in translational mode, where intensity reduction and broadening highlight structural relaxation. There is one peak nearly 280 $cm^{-1}$ corresponding to the translational mode of MgO-H and calcite impurity, and the peak intensity is lower and lower in other phlogopite samples, implying the bond breaking. At ~400 $cm^{-1}$, the translational mode of Mg-O vibrations combined with cationic modes shows notable weakening and broadening as exfoliation advances, reflecting increased strain and disorder. The ~700 $cm^{-1}$ peak, corresponding to Si(Al)-O bending vibrations, also undergoes a marked decline in intensity and peak sharpness, suggesting a disruption of the layered structure with prolonged exfoliation. [20-21]. For 2D phlogopite, the intensities of peaks associated with translational modes, such as K-O (~100 $cm^{-1}$) and O-H (~200 $cm^{-1}$), are drastically reduced, indicating weakened interlayer interactions. Additionally, the broadening and intensity reduction of the Si(Al)-O (~700 $cm^{-1}$) peak suggest substantial disruption in the layered structure with prolonged exfoliation.

Optical images of bulk and 2D phlogopite are shown in **Figure 2**a(i) and **2**a(ii), respectively. The optical image of bulk phlogopite (**Figure 2**a(i)) shows a highly reflective and irregularly shaped structure, and the edges of the bulk fragment appear sharp and crystalline, which is typical for unexfoliated mica, while that of 2D phlogopite (**Figure 2**a(ii)) shows flakes with a noticeable interference effect (pinkish-blue hues).

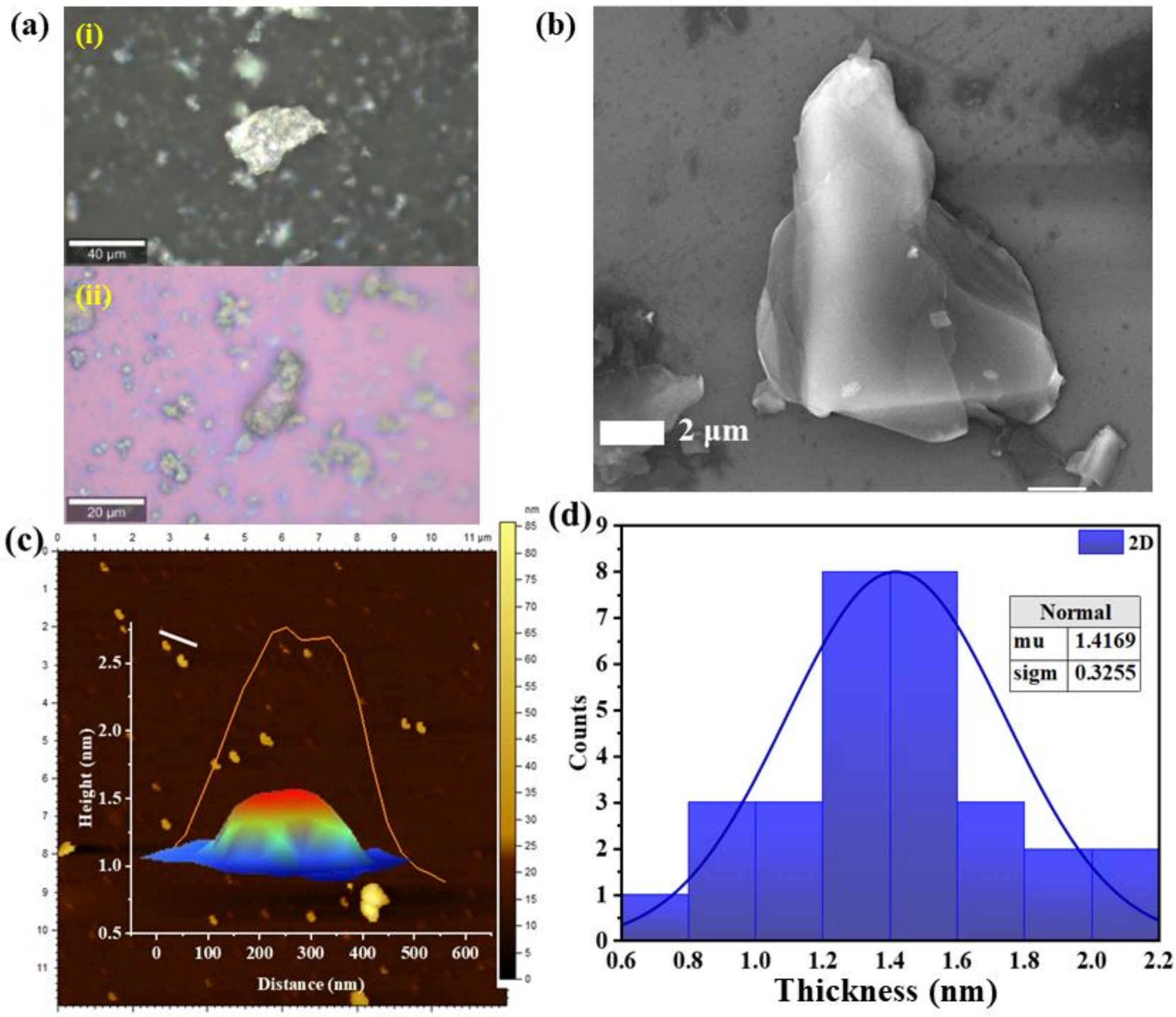

**Figure 2** (a) Optical images of (i) bulk, and (ii) exfoliated phlogopite, (b) SEM image of 4 h exfoliated phlogopite, (c) AFM image and line profile analysis of 2D phlogopite, (d) Thickness histogram of 2D phlogopite.

**Figure 2**b shows the SEM image of 4 h exfoliated phlogopite, clearly showing the layered structure. The corresponding EDAX analysis revealed material composition, which is shown in **Table 1**. In order to investigate the thickness of 2D sheets in the synthesized sample, AFM analysis was carried out, as shown in **Figure 2**c. The corresponding thickness histogram was obtained, as shown in **Figure 2**d. The average lateral width of the 2D phlogopite was found to be 479 nm, while the range of distribution is between 350 and 600 nm. The average particle

thickness measured was approximately 1.42 nm, with a distribution ranging between 0.5 and 2.5 nm.

**Table 1** Composition of phlogopite obtained using EDAX analysis.

| Elements | Mass% | Atom% |
|---|---|---|
| O | 43.12 | 57.10 |
| Mg | 2.46 | 2.14 |
| Al | 1.10 | 0.86 |
| Si | 51.78 | 39.06 |
| K | 1.55 | 0.48 |

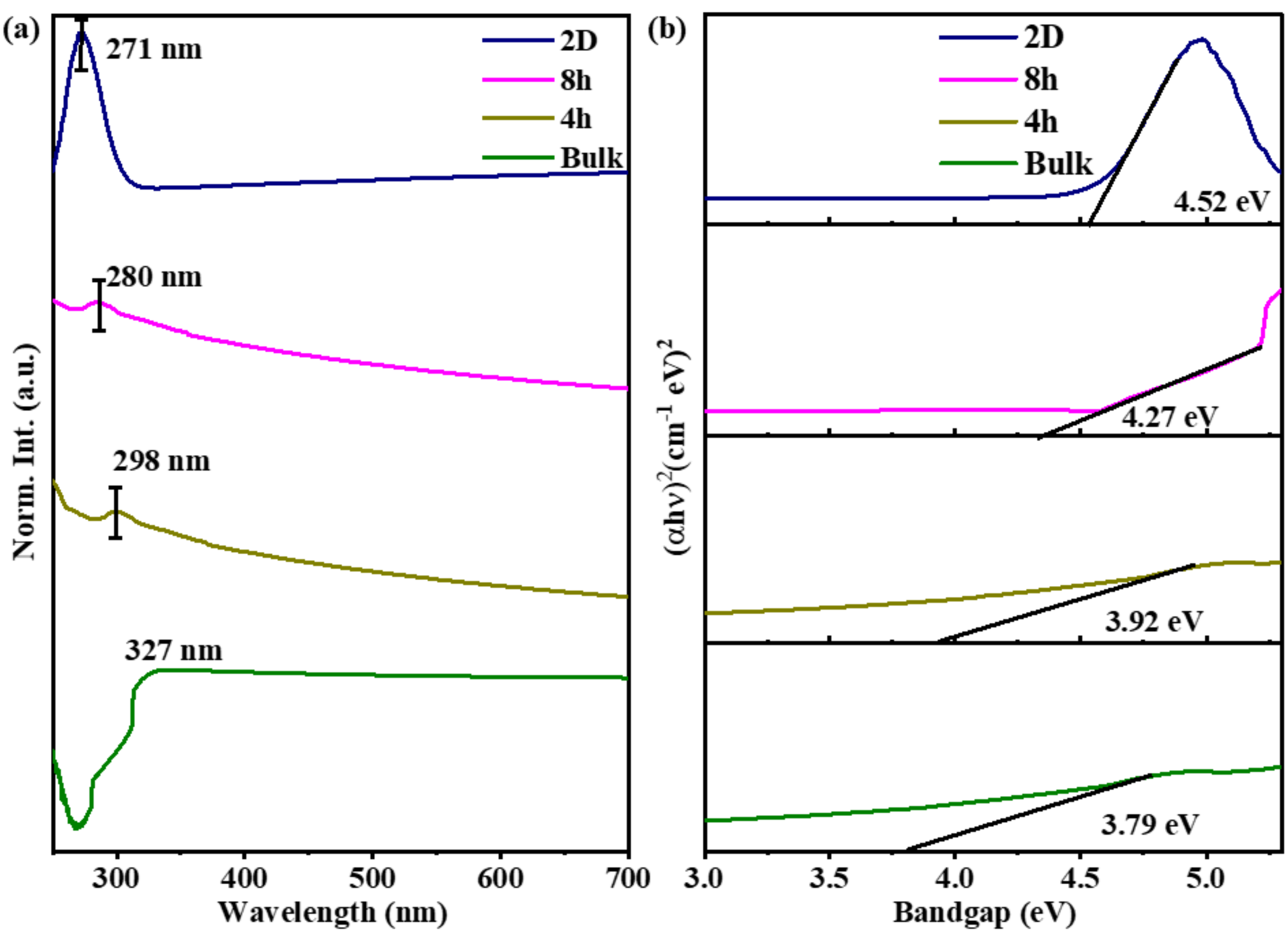

**Figure 3** (a) UV-Vis spectra of bulk, 4 h, 8 h, and 2D-phlogopite. (b) Corresponding Tauc plots for bulk, 4 h, 8 h, and 2D-phlogopite, along with calculated optical bandgaps.

To detect the changes in the optical bandgap, UV-Vis spectra were recorded at multiple stages during the exfoliation process, as shown in **Figure 3**a. The absorbance peak shifts from 327 nm for bulk phlogopite to 298 nm after 4 h of exfoliation, further to 280 nm after 8 h of exfoliation, and finally to 271 nm after the optimized 2D synthesis process is complete. The corresponding Tauc plots are shown in **Figure 3**b. A clear increase in the optical band gap from 3.79 eV for bulk phlogopite to 4.52 eV for 2D phlogopite is observed. This is due to the material's dimensionality being reduced significantly from bulk to the 2D limit, resulting in charge carriers being confined in the reduced dimensionality. The systematic increase in optical bandgap with prolonged exfoliation time (4 h vs. 8 h) further corroborates the reduction in layer thickness, consistent with the absorbance peak shift observed in the corresponding absorption spectra, as shown in **Figure 3**a.

### 3.2 Ultrafast Nonlinear Optical (NLO) Properties

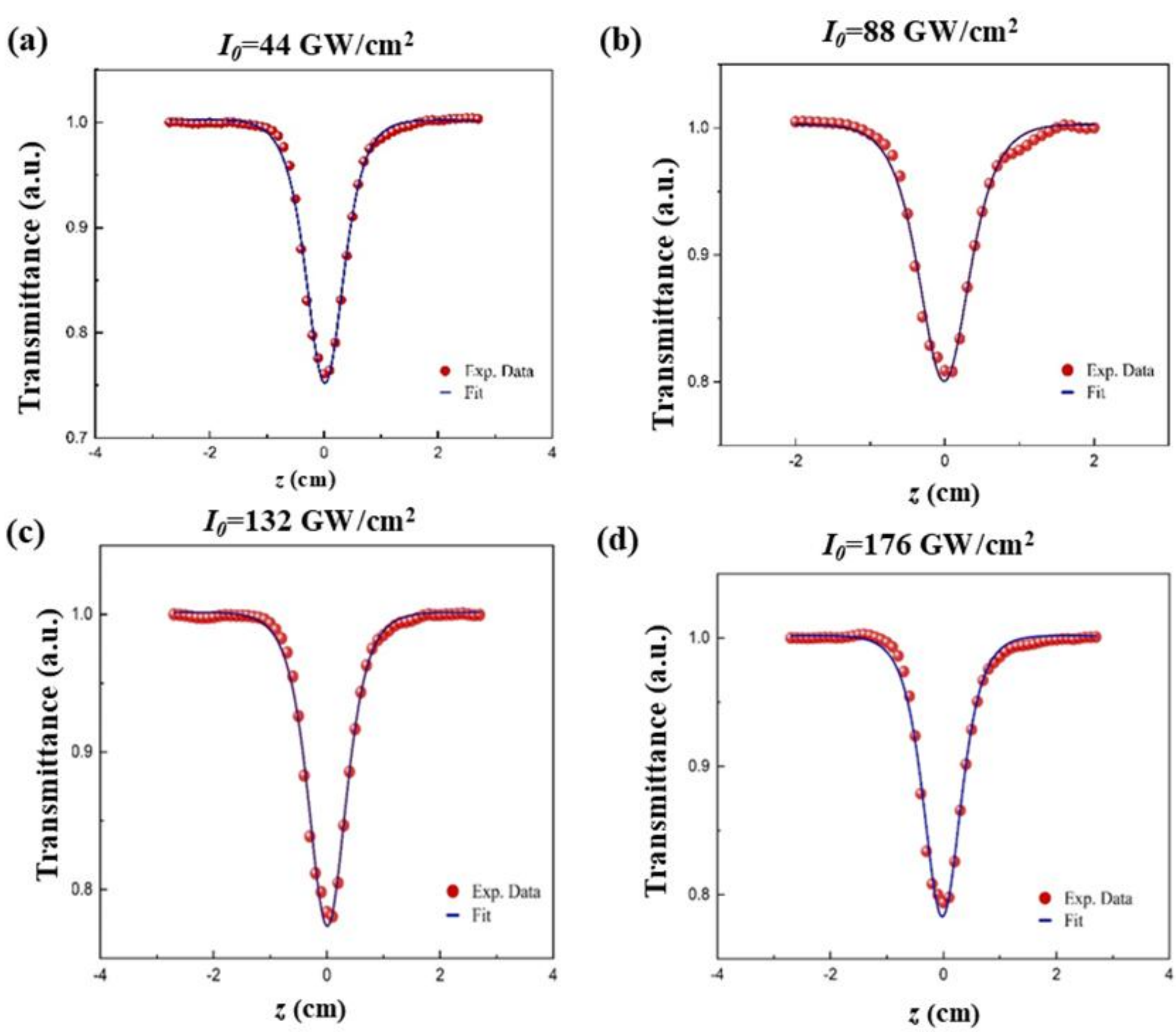

**Figure 4** Open aperture Z-scan data of 2D phlogopite at various intensities: (a) 44 GW/cm$^2$, (b) 88 GW/cm$^2$, (c) 132 GW/cm$^2$, and (d) 176 GW/cm$^2$.

The three-photon absorption behavior of 2D phlogopite was investigated using the open-aperture (OA) Z-scan technique. OA Z-scan data of 2D phlogopite at various intensities are shown in **Figure 4**a-d.

The sample was irradiated by a 100-fs pulse-width Ti:Sapphire laser (808nm) operating at 1 kHz repetition rate. In the Z-scan technique, the sample is translated through the laser beam's waist. The transmitted intensity passing through the sample can be calculated using the equation,

$$\frac{dI}{dz} = -\alpha I - \beta I^2 - \gamma I^3$$

**Equation 1**

Where intensity $I$ is calculated using the equation,

$$I(z,r) = I_0 \frac{\omega_0^2}{\omega^2(z)} \exp\left(-\frac{t}{\tau_p}\right) \exp\left(-\frac{2r^2}{\omega^2}\right)$$

**Equation 2**

Where ω is calculated using the equation,

$$\omega(z) = \omega_0 \sqrt[2]{(1 + z^2/z_r^2)}$$

**Equation 3**

Where α, β, and γ are linear, two-photon, and three-photon absorption coefficients, respectively. $\omega_0$, $z_r$, and $\tau_p$ are the beam waist, Rayleigh range, and pulse width, respectively. The experimental data were analysed based on a three-photon absorption (3PA) process, as the band gap of approximately 4.5 eV can be spanned by three photons from the 808 nm laser (each at 1.53 eV). The data were best fitted by a model incorporating both three-photon absorption (3PA) as the dominant mechanism and a secondary contribution from two-photon absorption (2PA). The data were best fitted by a model incorporating both three-photon absorption (3PA) as the dominant mechanism and a secondary contribution from two-photon absorption (2PA**).** The 3PA-2PA transmittance T can be given by,

$$T(z) = 1 - \frac{\beta I_0 L_{eff}}{2\sqrt{2}\left(1 + \frac{z^2}{z_r}\right)} - \frac{\gamma I_0^2 L_{eff}}{3\sqrt{3}\left(1 + \frac{z^2}{z_r^2}\right)^2}$$

**Equation 4**

Where $L_{eff}$ is calculated using the equation,

$$L_{eff} = \frac{1 - e^{-\alpha L}}{\alpha}$$

**Equation 5**

Where $L$ is the thickness of the sample ($L$= 2mm), and α is the linear absorption coefficient. The OA Z-scan experimental data, shown in **Figure 4**a-d, are fitted by **Equation 4**. The obtained values of the nonlinear absorption coefficient and cross-section are shown in **Table 2**. The three-photon absorption cross-section ($\sigma_{3PA}$) was found to be the highest ($16.86 \times 10^{-74}$ $cm^6$ $s^2/photon^2$), at an intensity of 44 $GW/cm^2$, and it decreases with further increases in intensity. Thus, absorption efficiency per photon decreases at higher power, which indicates saturable absorption (SA). SA is a phenomenon in which a material's light absorption diminishes as light intensity increases due to the depletion of the ground state of the absorbing particles. This is an intensity-dependent optical characteristic utilized in applications such as lasers, especially for the generation of brief pulses. [22] In 2D layered materials like phlogopite, at high photon flux, available states for multiphoton transitions get occupied. The effective population difference between ground and excited states decreases. Local heating due to high-intensity incident light changes the refractive index or absorption coefficient. All of these lead to reduced effective nonlinear absorption, consistent with three-photon saturable absorption (3PSA). [23]

**Table 2** Values of Optical Nonlinear Parameters of 2D phlogopite.

| **Pulse Energy (nJ)** | **Intensity** $\boldsymbol{I_0}$ $(\boldsymbol{GW/cm^2})$ | $\boldsymbol{\beta}$ $(\boldsymbol{10^{-3} cm/GW})$ | $\boldsymbol{\gamma}$ $(\boldsymbol{10^{-3} cm^3/GW^2})$ | $\boldsymbol{\sigma_{3PA}}$ $(\boldsymbol{10^{-74} cm^6.s^2/photon^2})$ |
| --- | --- | --- | --- | --- |
| 100 | 44 | 16.76 | 4.024 | 16.86 |
| 200 | 88.2 | 8.27 | 0. 840 | 3.50 |
| 300 | 132.2 | 4.49 | 0.394 | 1.65 |
| 400 | 176.4 | 4.89 | 0.216 | 0.91 |

### 3.3 Electrical Response from 2D Phlogopite

Electrical behavior under illumination is often the key to understanding the saturable process in semiconducting 2D materials like phlogopite. **Figure 5**a and **5**b show the electrical response from 2D phlogopite under different incident laser wavelengths of 532 nm and 650 nm, respectively, at a constant voltage of 5V. The current decreases with increasing intensity of the incident light. Hence, optically, the material seems to saturate, and electrically, it shows reduced conductivity at higher light intensities. At higher photon flux, trap states become saturated due to lattice defects, edges, or impurities. Once traps are filled, recombination becomes dominant, reducing free carrier density. This leads to lower photocurrent despite higher illumination. This phenomenon is common in layered silicates and mica-type 2D materials. [24-26] At intensities of tens of GW/cm², the effects of local heating need to be considered. Increased temperature enhances phonon scattering (reducing mobility) and modifies the bandgap slightly (reducing absorption). This manifests as decreasing current with higher laser power. Moreover, since our optical measurements already show SA, the number of absorbed photons per unit intensity increase drops beyond the saturation point. Fewer additional carriers are generated, so the photocurrent plateaus or decreases. This is consistent with our observed decrease in $\sigma_{3PA}$ and photocurrent at high intensities. To summarise, the system transitions from a low-intensity regime (efficient multiphoton absorption, generating carriers) to an intermediate regime (~44 GW/cm$^2$) (maximum absorption, maximum carrier generation), and finally to a high-intensity regime (state filling, recombination, and possible thermal effects), which manifests as a decrease in both absorption and photocurrent. Thus, 2D phlogopite exhibits 3PSA, coupled with reverse photoconductivity, both of which are caused by carrier trapping and state saturation at high excitation intensities.

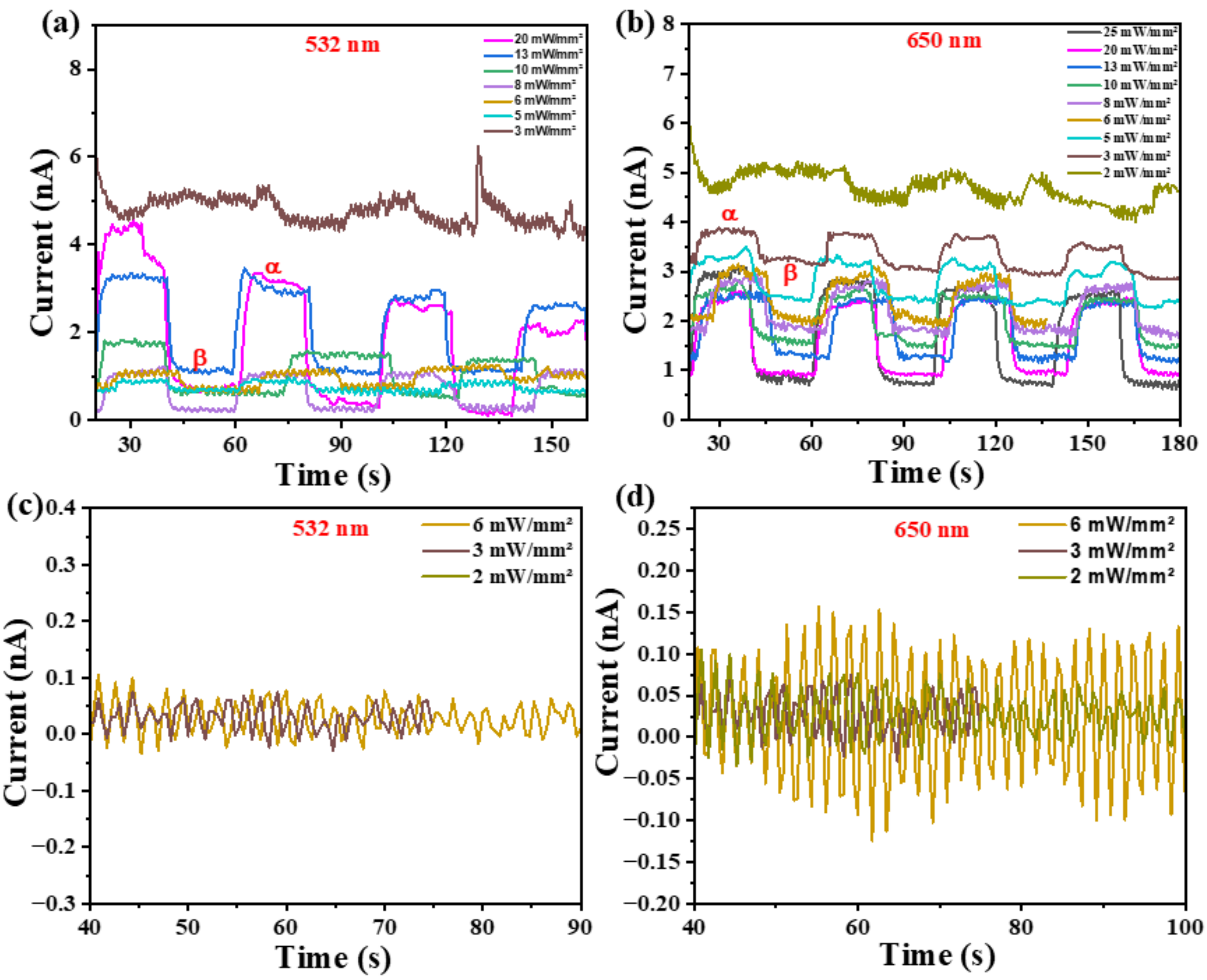

**Figure 5** (a) Current as a function of time for 2D phlogopite under 532 nm illumination at varying light intensities, with α representing the light-off condition and β the light-on condition, (b) Current as a function of time for 2D phlogopite under 650 nm illumination at varying light intensities, (c) Time-resolved current under 532 nm illumination, and (d) time-resolved current under 650 nm illumination for bulk phlogopite at different light intensities, all in constant voltage of 5V.

It has been demonstrated that the bulk phlogopite (**Figure 5**c for 532 nm and **Figure 5**d for 650 nm) does not show SA-like behavior, as it did not respond efficiently to the laser light. The insensibility to laser light explains changes in structural integrity or the electronic property of the material during prolonged exfoliation, which affects the mechanisms through which excited-state absorption happens. So, it becomes clear that the extent of exfoliation determines whether the material will perform according to its optical properties. Although 2D phlogopite that was exfoliated for longer periods (~ 6 h) shows SA behavior, the bulk diminishes that ability because of the interference of the layered structure, as the interaction with the laser is reduced.

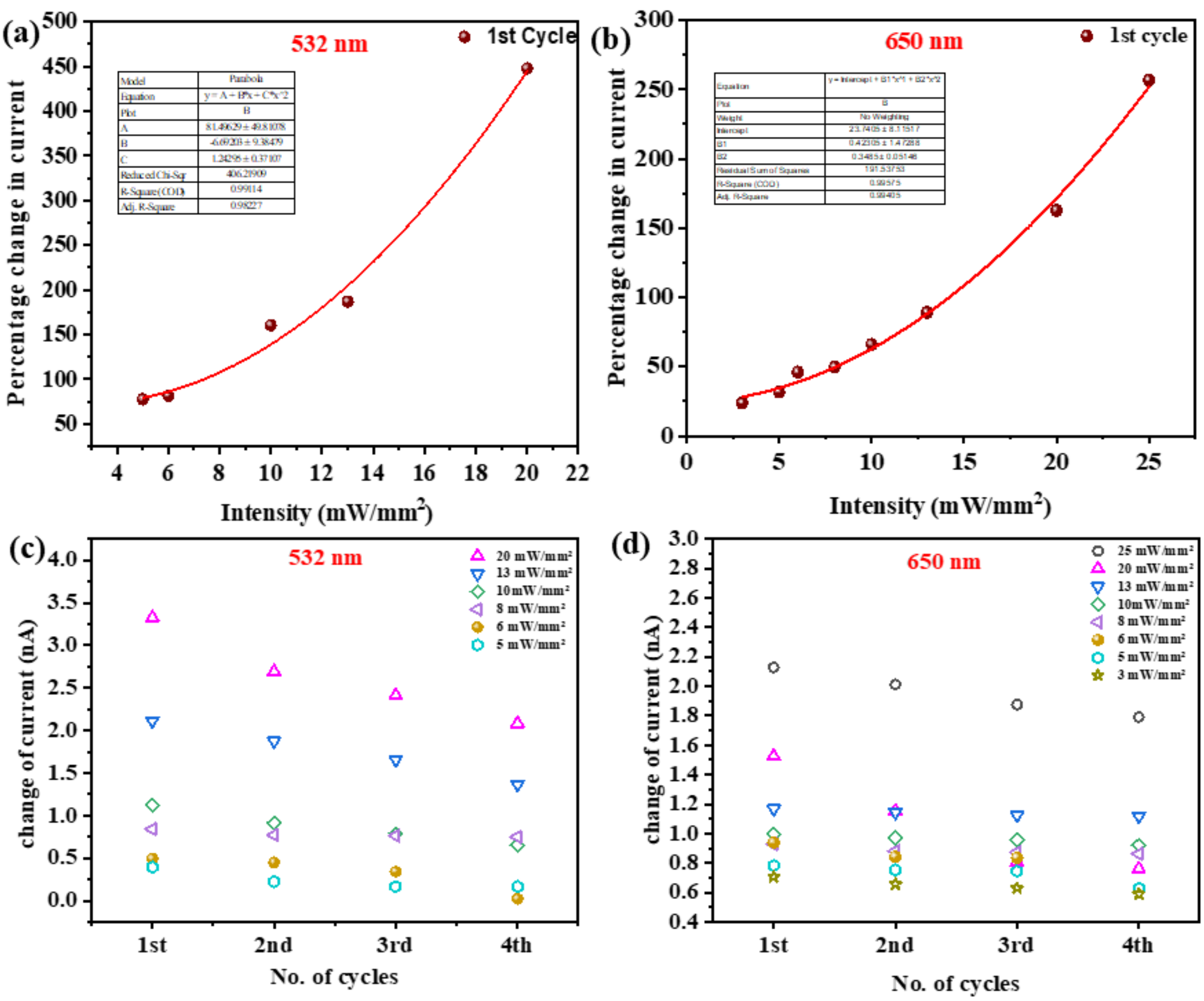

**Figure 6** (a) Percentage change in current with an increase in intensity under (a) 532 nm and (b) 650 nm laser incidence. Change in current tracked for repeated cycles under (a) 532 nm and (b) 650 nm laser incidence.

To further the discussion on the recombination kinetics and carrier dynamics, the percentage change in photocurrent was plotted as a function of incident laser intensity for both 532 nm and 650 nm wavelengths, as shown in **Figure 6**a-b. As illustrated in the plots, the electrical response of the 2D phlogopite device does not follow a simple linear dependency on photon flux. Instead, the percentage change in current exhibits a distinct superlinear (parabolic) growth with increasing laser intensity. The experimental data for both wavelengths were fitted with a second-order polynomial model ($y = A + Bx + Cx^2$), yielding high coefficients of determination ($R^2$ is approx 0.99), which confirms the nonlinear nature of the photo-induced modulation. This parabolic trajectory indicates that the efficiency of the resistive switching (or current quenching) enhances significantly at higher optical power densities. In the context of the previously observed negative photoconductivity, this behavior can be attributed to the progressive dominance of nonlinear optical absorption and trap-assisted recombination mechanisms. At lower intensities (< 5 $mW/mm^2$), the change in current is moderate, governed primarily by linear absorption and initial trap filling. However, as the intensity increases towards 20 $mW/mm^2$, the response accelerates. This suggests that the generation of photo-excited carriers, and their subsequent trapping or recombination, is scaling nonlinearly, likely driven by multi-photon absorption processes (such as 2PA and 3PA) which are known to scale with the square or cube of the intensity ($I^2$ or $I^3$). Furthermore, mathematically, as the incident light drives the device current closer to zero (deep depletion state observed in the time-resolved scans), the relative percentage change increases asymptotically. This strong nonlinear dependence confirms that 2D phlogopite acts as a highly sensitive, optically-gated variable resistor, where the "off-state" depth can be tuned nonlinearly via laser intensity.

To evaluate the dynamic reliability of the 2D phlogopite photodetector, the current modulation depth was monitored over successive switching cycles. **Figure 6**c-d presents the variation in current change as a function of cycle number for incident wavelengths of 532 nm and 650 nm. It has been observed that the amplitude of the photoreponse is not a constant but decreases monotonically from cycle to cycle when optically switched. For example, at an intensity of 20 $mW/mm^2$ (532 nm), the increase in current is about 3.3 nA on the first cycle and decreases to around 2.1 nA on the fourth cycle. A similar downward trend is evident for all of the intensity and wavelength combinations tested. This degradation in photoresponse from one cycle to the next is often referred to as optical fatigue. This optical fatigue indicates that, as time passes, the effectiveness of the device decreases. Because the photoresponse mechanism involves carrier trapping, the instability of the photoresponse indicates that trap states do not completely recover during the "light off" portions of the cycles.

Deep-level defects have a physical origin that can both trap carriers and reduce current associated with them; they are characterised by a specific time period in which they delay the release or detrapping of carriers from the defect upon removal of the excitation energy (light). If the rate of a system's switching frequency exceeds the rate of slow detrapping kinetics, some portion of trap sites will still be occupied at the beginning of the following cycle. As a consequence, there are fewer trap sites available to trap the free carriers in the subsequent cycle, thus resulting in decreasing reductions in current. This memory effect highlights that although this material is highly responsive, the time it takes to recover from the material's defect site is related to the rate at which charge persists on defect sites.

### 3.3 Theoretical Simulation

The atomic models describing the K-Mg-Si-O system in phlogopite are shown in **Figure 5**a-b. In **Figure 5**a, multiple atomic layers are periodically staggered, forming a three-dimensional lattice with strong interlayer bonding, ensuring high structural stability. This bulk arrangement has strong electronic properties coupled with mechanical toughness, and it is ideal for general use that demands intrinsic stability. However, upon transitioning to the 2D structure, as shown in **Figure 5**b, it consists of a surface-terminated layer with reduced coordination of atoms, leading to weakened interlayer interactions and enhanced surface activity. This dimensional reduction can induce band gap widening, altered charge transport properties, and enhanced reactivity, which are desirable traits for nanoelectronics, optoelectronics, and catalytic applications.

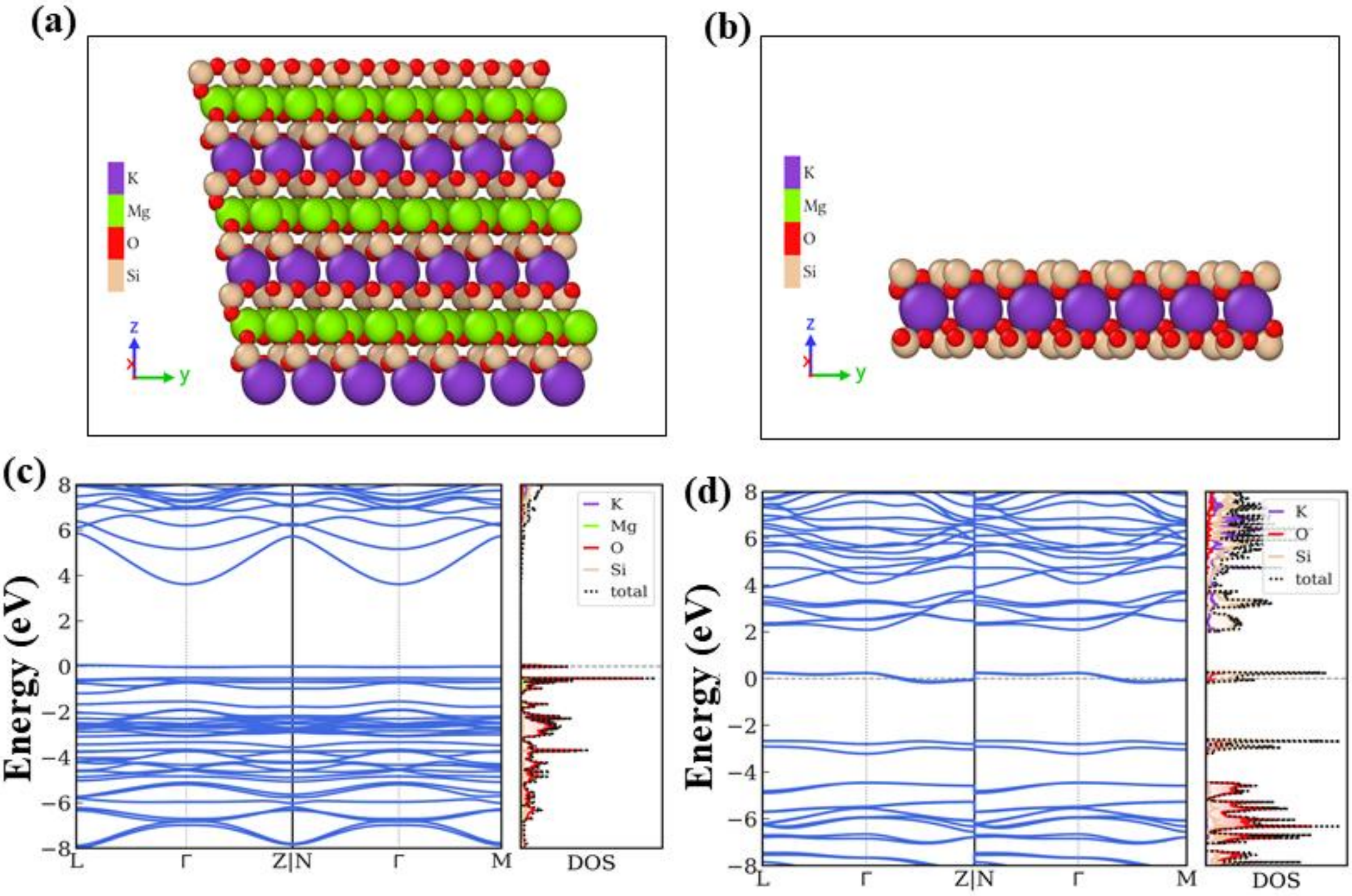

**Figure 5** (a) Atomic configuration of the bulk structure, (b) Corresponding 2D monolayer structure obtained by exfoliation, highlighting the reduced dimensionality and enhanced surface exposure, (c) Electronic band structure and density of states (DOS) of the bulk system, (d) Electronic band structure and DOS of the 2D structure.

**Figure 5**c corresponds to the band structure of bulk phlogopite, which has a band gap of 3.8 eV, indicating its semi-conducting behavior. The density of states (DOS) analysis suggests that oxygen and silicon contribute significantly to states near the valence and conduction bands. **Figure 5**d demonstrates the band structure of its corresponding 2D structure, which has a band gap of 4.5 eV. The energy state near the Fermi level is the intermediate energy state. The DOS plot highlights significant contributions from potassium, oxygen, and silicon, with energy-dependent distributions affecting electronic characteristics.

The theoretical band structure and DOS analysis (**Figure 5**c-d) reveal that the bandgap of 2D phlogopite widens to approximately 4.5 eV compared to 3.8 eV in bulk, while intermediate states emerge near the Fermi level due to defects, impurities, and surface effects. This is consistent with our experimental findings, as shown in **Figure 3**b. These intermediate states facilitate stepwise multiphoton transitions from the valence band to the conduction band, enabling efficient 3PA under high-intensity illumination. At moderate intensities (~ 44 GW/cm²), resonant coupling through these states enhances the 3PA cross-section, while state filling and carrier recombination at higher intensities lead to a reduction in absorption, which manifests as 3PSA. This interplay between bandgap widening and defect-assisted intermediate states provides a consistent theoretical explanation for the observed nonlinear optical and electrical responses in 2D phlogopite.

## 4. Conclusions

In summary, the present work demonstrated successful fabrication of 2D phlogopite by the LPE method, which thus confirms its effectiveness for accessing high-quality, atomically thin layers. A deep analysis by structural, optical, and electrical characterizations sheds light on the distinctive characteristics of 2D phlogopite, namely its large band gap, optical response, and exceptionally high stability. XRD confirmation indicated preferential (033) plane orientation; however, UV-Vis indicated the tuned band gap at varied exfoliation times, with high stability and layer dependence. Raman spectroscopy reveals significant structural and compositional changes. Morphological investigations through AFM and SEM strongly affirm the homogeneity and nanoscale thickness and reveal the silicate-rich nature of the exfoliated layers. The observed SA demonstrated a promising optical limiting and switching effect as well as various applications for 2D phlogopite, most significantly for photonic devices; however, the most stable SA was confirmed when optimised conditions for exfoliation were implemented. And these properties strongly emphasised the potential applications of 2D phlogopite in photonics and optoelectronics, facilitating its incorporation into high-intensity light applications, such as optical limiting, switching, and mode-locking.

## CRediT authorship contribution statement

**Nabarun Mandal, Sagnik Chakraborty**: Conceptualization, Methodology, Formal Analysis, Investigation, Writing Original Draft, Visualization, Editing, and Reviewing; **Ranjeet Singh, Jhionathan de Lima, Saswata Goswami, Binay Bhushan**: Methodology and Investigation; **Rahul Rao, Nicholas Glavin, Ajit K Roy, Vidya Kochat, Cristiano Francisco Woellner, Prasanta Kumar Datta**: Reviewing and Validation; **Chandra Sekhar Tiwary**: Supervision, Project administration, Resources, Reviewing, and Validation.

## Declaration of Competing Interest

The authors declare that they have no known competing financial interests or personal relationships that could have appeared to influence the work reported in this paper.

## Acknowledgments

The authors acknowledge the Indian Institute of Technology Kharagpur (IITKGP) for laboratory facility support.

## Graphical Abstract

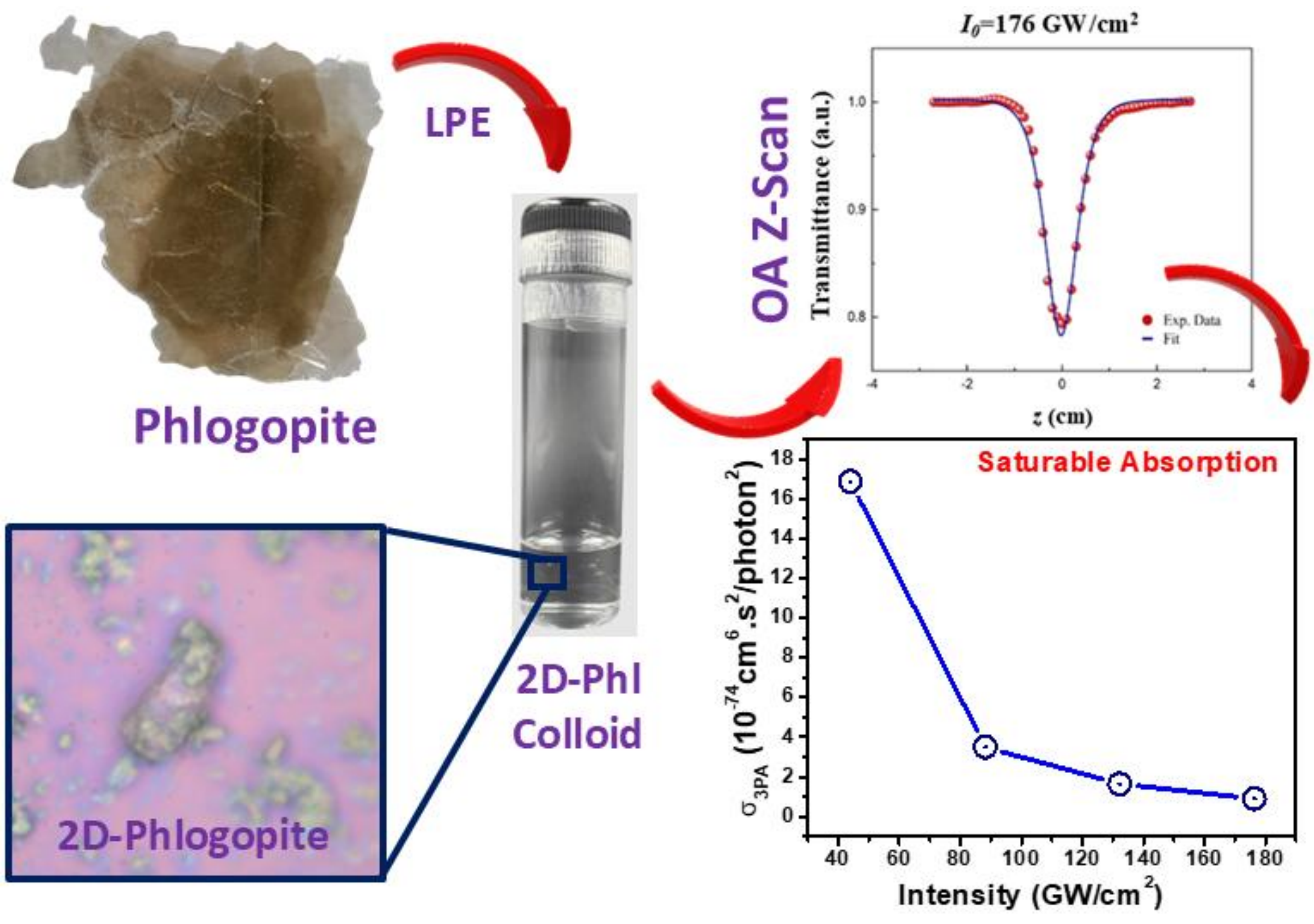